\documentclass[aps,prb,twocolumn,superscriptaddress,showpacs,showkeys,amsmath,amssymb]{revtex4}

\usepackage{epsfig,amsmath,amssymb,color}
\renewcommand{\theequation}{\arabic{section}.\arabic{equation}}

\begin{document}

\title{Frustrated honeycomb-lattice bilayer quantum antiferromagnet in a magnetic field: 
       Unconventional phase transitions in a two-dimensional isotropic Heisenberg model}

\author{Taras Krokhmalskii}
\affiliation{Institute for Condensed Matter Physics,
          National Academy of Sciences of Ukraine,
          Svientsitskii Street 1, 79011 L'viv, Ukraine}
\affiliation{Department for Theoretical Physics,
          Ivan Franko National University of L'viv,
          Drahomanov Street 12, 79005 L'viv, Ukraine}

\author{Vasyl Baliha}
\affiliation{Institute for Condensed Matter Physics,
          National Academy of Sciences of Ukraine,
          Svientsitskii Street 1, 79011 L'viv, Ukraine}

\author{Oleg Derzhko}
\affiliation{Institute for Condensed Matter Physics,
          National Academy of Sciences of Ukraine,
          Svientsitskii Street 1, 79011 L'viv, Ukraine}
\affiliation{Institut f\"{u}r theoretische Physik,
          Otto-von-Guericke-Universit\"{a}t Magdeburg,
          P.O. Box 4120, 39016 Magdeburg, Germany}
\affiliation{Department for Theoretical Physics,
          Ivan Franko National University of L'viv,
          Drahomanov Street 12, 79005 L'viv, Ukraine}
\affiliation{Abdus Salam International Centre for Theoretical Physics,
          Strada Costiera 11, 34151 Trieste, Italy}

\author{J\"{o}rg Schulenburg}
\affiliation{Universit\"{a}tsrechenzentrum,
          Otto-von-Guericke-Universit\"{a}t Magdeburg,
          P.O. Box 4120, 39016 Magdeburg, Germany}

\author{Johannes Richter}
\affiliation{Institut f\"{u}r theoretische Physik,
          Otto-von-Guericke-Universit\"{a}t Magdeburg,
          P.O. Box 4120, 39016 Magdeburg, Germany}

\date{\today}



\begin{abstract}
We consider the spin-1/2 antiferromagnetic Heisenberg model on a bilayer honeycomb lattice including interlayer frustration
in the presence of an external magnetic field.
In the vicinity of the saturation field,
we map the low-energy states of this quantum system onto the spatial configurations of hard hexagons on a honeycomb lattice.
As a result,
we can construct effective classical models (lattice-gas as well as Ising models) on the honeycomb lattice
to calculate the properties of the frustrated quantum Heisenberg spin system in the low-temperature regime.
We perform classical Monte Carlo simulations for a hard-hexagon model and adopt known results for an Ising model
to discuss the finite-temperature order-disorder phase transition
that is driven by a magnetic field at low temperatures.
We also discuss an effective-model description around the ideal frustration case 
and find indications for a spin-flop like transition in the considered isotropic spin model.
\end{abstract}

\maketitle

\section{Introduction}
\label{sec1}
\setcounter{equation}{0}

An important class of quantum Heisenberg antiferromagnets consists of the so-called two-dimensional dimerized quantum antiferromagnets.
They can be obtained by placing strongly antiferromagnetically interacting pairs of spins 1/2 (dimers) on a regular two-dimensional lattice 
and assuming weak antiferromagnetic interactions between dimers.
Among such models one may mention the $J-J^\prime$ model
with the staggered arrangement of the strong $J^\prime$ bonds 
(defining dimers and favoring singlet formation on dimers)
on a square lattice\cite{jjprime}
(see also Ref.~\onlinecite{valenti-eggert} for related dimerized square-lattice models).
Other examples are the bilayer models:
They consist of two antiferromagnets in each layer with a dominant nearest-neighbor interlayer coupling which defines dimers.\cite{bilayer}
By considering additional frustrating interlayer couplings 
the bilayer model can be pushed in the parameter space to a point 
which admits a rather comprehensive analysis of the energy spectrum.\cite{andreas}
For this special set of coupling parameters, the frustrated bilayer is a system with local conservation laws
(the square of the total spin of each dimer is a good quantum number)
that explains why it is much easier to examine this specific case.
On the other hand,
the frustrated bilayer belongs to the class of so-called localized-magnon spin systems,\cite{prl2002}
which exhibit some prominent features around the saturation field, 
such as 
a ground-state magnetization jump at the saturation field,
a finite residual entropy at the saturation field,
and 
an unconventional low-temperature thermodynamics, 
for a review see Refs.~\onlinecite{review1,review2,review3}.
The singlet state of the dimer is the localized-magnon state which belongs to a completely dispersionless (flat) one-magnon band.
Over the last decade  a large  variety of flat-band systems with unconventional physical properties was found, 
see Refs.~\onlinecite{other1,other2,other3} and references therein.
For the flat-band systems at hand, 
the local nature of the one-magnon states allows to construct also localized many-magnon states 
and to calculate their degeneracy by mapping the problem onto a classical hard-core-object lattice gas;
the case of the frustrated bilayer was discussed in Refs.~\onlinecite{prb2006,prb2010}.
In the strong-field low-temperature regime
the independent localized-magnon states are the lowest-energy ones 
and therefore they dominate the thermodynamics.
The thermodynamic properties in this regime can be efficiently calculated using classical Monte Carlo simulations for a lattice-gas problem.
Even in case of small deviations from the ideal flat-band geometry 
a description which is based on the strong-coupling approach\cite{aa} can be elaborated.\cite{around} 
Again the effective theory is much simpler than that for the initial problem. 

From the theoretical side, frustrated bilayer systems have been studied by several authors.
Thus,
the frustrated square-lattice bilayer quantum Heisenberg antiferromagnet was studied in Refs.~\onlinecite{lin,prb2006,prb2010,chen,albuquerque,murakami,alet},
whereas the honeycomb-lattice bilayer with frustration was studied in 
Refs.~\onlinecite{oitmaa,zhang,bishop} (intralayer frustration) 
and 
Refs.~\onlinecite{classical,brenig} (interlayer frustration).
For the system to be examined in our paper,
i.e., 
the spin-1/2 antiferromagnetic Heisenberg model on a bilayer honeycomb lattice including interlayer frustration,
H.~Zhang et al.\cite{brenig} have determined the quantum phase diagram at zero magnetic field 
for a rather general case of an arbitrary relation between the nearest-neighbor intralayer coupling and the frustrating interlayer coupling.
Another recent study reported in Ref.~\onlinecite{classical} 
concerns the antiferromagnetic classical Heisenberg model on a bilayer honeycomb lattice in a highly frustrated regime 
in the presence of a magnetic field.
Its main result is the phase diagram of the model in the plane ``magnetic field -- temperature''.
However, this analysis cannot contain any hallmarks caused by the localized magnons, 
since localized-magnon features represent a pure quantum effect which disappears in the classical limit.

From the experimental side,
one may mention several layered materials, which can be viewed as frustrated bilayer quantum Heisenberg antiferromagnets.
Thus, 
the compound Ba$_2$CoSi$_2$O$_6$Cl$_2$ could be described as a two-dimensionally antiferromagnetically coupled spin-1/2 $XY$-like spin dimer system
in which Co$^{2+}$ sites form the frustrated square-lattice bilayer.\cite{tanaka}
The interest in the frustrated honeycomb-lattice bilayers stems from experiments on Bi$_3$Mn$_4$O$_{12}$(NO$_3$).\cite{smirnova}
In this compound,
the ions Mn$^{4+}$ form a frustrated spin-$3/2$ bilayer honeycomb lattice.\cite{kandpal}
Finally, let us mention that a bilayer honeycomb lattice can be realized using ultracold atoms.\cite{cold_atoms}

The present study has several goals.
Motivated by the recent paper of H.~Zhang et al.,\cite{brenig} we wish to extend it
to the case of nonzero magnetic field.
On the other hand, with our study we complement the analysis of the classical case\cite{classical} to the pure quantum case of $s=1/2$.
Finally, 
the present study can be viewed as an extension of our previous calculations\cite{prb2006,prb2010} to the honeycomb-lattice geometry.
Although we do not intend to provide a theoretical description of Bi$_3$Mn$_4$O$_{12}$(NO$_3$),
our results may be relevant for the discussion of the localized-magnon effects in this and in similar materials.

The outline of the paper is as follows.
Section~\ref{sec2} contains the spectroscopic study of the frustrated honeycomb-lattice bilayer spin-1/2 Heisenberg antiferromagnet:
By exact diagonalization for finite quantum systems and direct calculations for finite hard-core lattice-gas systems
we show the correspondence between the ground states in the large-$S^z$ subspaces and the spatial configurations of hard hexagons on an auxiliary honeycomb lattice.
Based on the established correspondence,
in Section~\ref{sec3}
we report results of classical Monte Carlo simulations for hard hexagons on the honeycomb lattice
and use them to predict the properties of the frustrated honeycomb-lattice bilayer spin-1/2 Heisenberg antiferromagnet 
in the strong-field low-temperature regime.
The most intriguing outcome is an order-disorder phase transition which is expected at low temperatures just below the saturation
field.
This transition  is related to the ordering of the localized magnons on the two-sublattice honeycomb lattice as the density of the localized magnons increases.
Section~\ref{sec4} deals with some generalization of the independent localized-magnon picture:
We show how to take into account the contribution of a low-lying set of other localized states
as well as discuss the effect of deviations from the ideal frustration
case.
We end with a summarizing discussion in Section~\ref{sec5}.
Several technical details are put to the appendixes.

\section{Independent localized-magnon states}
\label{sec2}
\setcounter{equation}{0}

\begin{figure}
\begin{center}
\includegraphics[clip=on,width=80mm,angle=0]{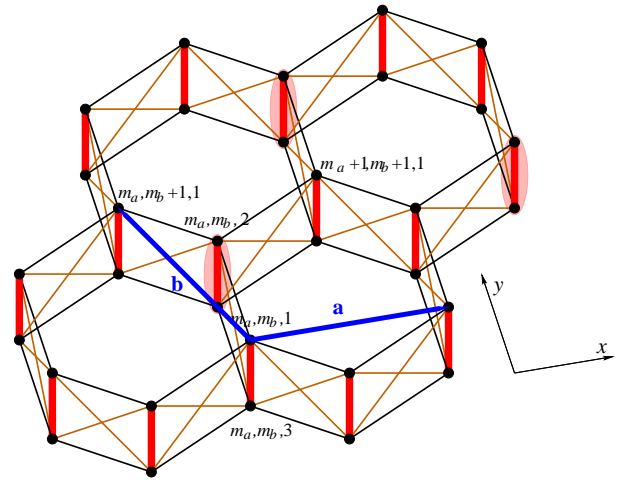}
\caption{(Color online)
Frustrated honeycomb-lattice bilayer.
It can be considered as a triangular lattice with four sites in the unit cell.
${\bf{a}}$ and ${\bf{b}}$ are the basis vectors for the triangular lattice
and the integer numbers $m_a$ and $m_b$ determine the position of the unit cell.
The vertical (red) bonds have the strength $J_2$.
The nearest-neighbor intralayer (black) bonds have the strength $J_1$.
The frustrating interlayer (brown) bonds have the strength $J_{\mbox{x}}$.
The main focus of our study is the case $J_{\mbox{x}}=J_1$, $J_2>3J_1$ (ideal
frustration case).
The case $J_{\mbox{x}}\ne J_1$, $\vert J_1-J_{\mbox{x}}\vert/J_2\ll 1$ is considered in Sec.~\ref{sec4b}.}
\label{f01}
\end{center}
\end{figure}

In the present paper,
we consider the spin-1/2 Heisenberg antiferromagnet with the Hamiltonian
\begin{eqnarray}
\label{201}
H=\sum_{\langle i j\rangle}J_{ij} {\bf{s}}_i\cdot {\bf{s}}_j -hS^z,
\;\;\;
J_{ij}>0,
\;\;\;
S^z=\sum_i s_i^z
\end{eqnarray}
defined on the honeycomb-lattice bilayer shown in Fig.~\ref{f01}.
The first sum in Eq.~(\ref{201}) runs over all bonds of the lattice 
and hence $J_{ij}$ acquires three values: 
$J_2$ (dimer bonds), 
$J_1$ (nearest-neighbor intralayer bonds), 
and 
$J_{\mbox{x}}$ (frustrating interlayer bonds), 
see Fig.~\ref{f01}.
In what follows 
we consider  the case $J_{\mbox{x}}=J_1$ and call it the ``ideal frustration case''
(or  ``ideal flat-band case'').
Only in Sec.~\ref{sec4b} we discuss deviations from the ideal frustration case, 
i.e., $J_{\mbox{x}}\ne J_1$.
Since the $z$ component of the total spin $S^z$ commutes with the Hamiltonian
we can consider the subspaces with different values of $S^z$ separately.

In the strong-field regime the subspaces with large $S^z$ are relevant.
The only state with $S^z=N/2$ is the fully polarized state
$\vert\ldots\uparrow\ldots\rangle$
with the energy
$E_{\rm{FM}}=N(J_2/8+3J_1/4)$.
In the subspace with $S^z=N/2-1$ (one-magnon subspace)
$N$ eigenstates of $H$ (\ref{201}) belong to four one-magnon bands,
$E_{\rm{FM}}+\Lambda_{{\bf{k}}}^{(\alpha)}$, $\alpha=1,2,3,4$,
with the dispersion relations:
\begin{eqnarray}
\label{202}
\Lambda_{{\bf{k}}}^{(1)}=\Lambda_{{\bf{k}}}^{(2)}
=-J_2-3J_1,
\;\;\;
\Lambda_{{\bf{k}}}^{(3,4)}
=-3J_1\mp J_1\vert \gamma_{{\bf{k}}} \vert,
\nonumber\\
\vert\gamma({\bf{k}})\vert
=
\sqrt{3+2\left[\cos k_a +\cos k_b +\cos\left(k_a+k_b\right)\right]}. \;\;
\end{eqnarray}
Here  
${\bf{k}}=(k_x,k_y)$,
$k_a=\sqrt{3}a_0 k_x$, 
$k_b=3a_0k_y/2-\sqrt{3}a_0k_x/2$,
where $a_0$ is the hexagon side length,
and ${\bf{k}}$ acquires ${\cal{N}}/2$ values from the first Brillouin zone, 
see Appendix~A.
The ${\cal{N}}$ states from the two flat bands $\alpha=1$ and $\alpha=2$ can be chosen as a set of localized states 
where the spin flip is located on one of the ${\cal{N}}$ vertical dimers,
see Fig.~\ref{f01}.
The remaining ${\cal{N}}$ states 
(i.e., from the two dispersive bands $\alpha=3$ and $\alpha=4$) 
are extended over the whole lattice.
As can be seen from Eq.~(\ref{202}),
the two-fold degenerate dispersionless (flat) one-magnon band becomes the lowest-energy one,
if $J_2> 3J_1$,
i.e., if the  strength of the vertical bond $J_2$ is sufficiently large.
In what follows, 
we assume that this inequality holds.
From the one-magnon spectra (\ref{202})
we can also get  the value of the saturation field:
$h_{\rm{sat}}=J_2+3J_1$.

\begin{figure}
\begin{center}
\includegraphics[clip=on,width=80mm,angle=0]{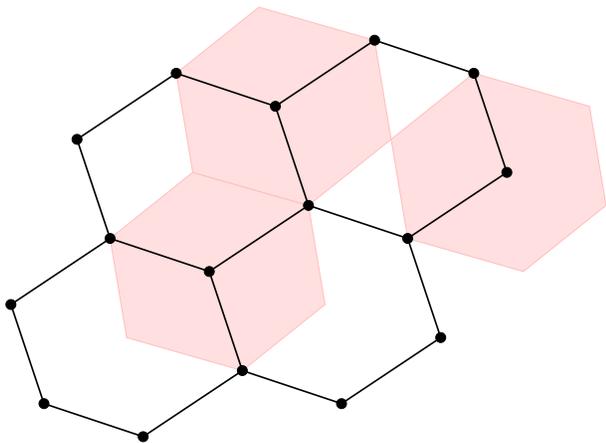}
\caption{(Color online)
Independent localized-magnon states
(corresponding to the shaded vertical dimers in Fig.~\ref{f01})
and hard-hexagon configurations on an auxiliary honeycomb lattice.}
\label{f02}
\end{center}
\end{figure}

We pass to the many-magnon ground states.
Because the localized one-magnon states have the lowest energy in the one-magnon subspace,
the ground states in the subspaces with $S^z=N/2-n$, $n=2,\ldots, n_{\max}$,
$n_{\max}={\cal{N}}/2=N/4$
can be obtained by populating the dimers.
However, for the ground-state manifold a hard-core constraint is valid,
i.e.,
the neighboring vertical dimers cannot be populated simultaneously, 
since the occupation of nearest neighbors leads to an increase of the energy.
Thus, we arrive at the mapping onto a classical lattice-gas model of hard hexagons on an auxiliary honeycomb lattice:
Each ground state of the quantum spin model can be visualized as a spatial configuration of the hard hexagons on the honeycomb lattice 
excluding the population of neighboring sites (hard-core rule),
see Fig.~\ref{f02}.

The occupation of neighboring sites, 
excluded for the ground-state manifold at $S^z=N/2-2,\ldots, N/4$,
provides another class of localized states 
which can be visualized as overlapping hexagons on the honeycomb lattice:
These states were completely characterized in Refs.~\onlinecite{andreas} and \onlinecite{prb2010}.
Each overlapping pair of hexagons 
(i.e., occupation of neighboring dimers by localized magnons) 
increases the energy by $J_1$.
If $J_2/J_1$ is sufficiently large,
the overlapping hexagon states are the lowest excited states in the subspaces with $S^z=N/2-2,\ldots,N/4$, 
but they are the ground states in the subspaces with lower $S^z$.
From exact-diagonalization data for $N=24,\,32,\,36,\,48$ 
we determined the required values of $J_2/J_1$ as 3.687, 3.781, 3.813, 3.874, respectively:
For these values the first excited state in the subspace with  $S^z=N/2-2$ are the overlapping hexagon states.

\begin{figure}
\begin{center}
\includegraphics[clip=on,width=85mm,angle=0]{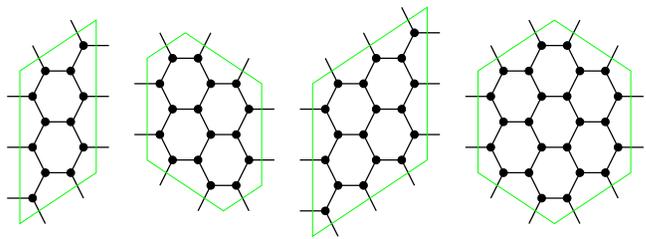}
\caption{(Color online)
Finite lattices used in our exact-diagonalization studies.
${\cal{N}}=12,\,16,\,18,\,24$, i.e. $N=24,\,32,\,36,\,48$ (from left to right).
Periodic boundary conditions are implied.}
\label{f03}
\end{center}
\end{figure}

We check our statements on the character of the ground states and the excited states by comparison with exact-diagonalization data.
Clearly,
exact diagonalizations are restricted to finite lattices,
which are shown in Fig.~\ref{f03}.
We use the {\it spinpack} package  \cite{spin} 
and exploit the local symmetries to perform numerical exact calculations for large sizes of the Hamiltonian matrix. 
The ground-state degeneracy coincides with the number of spatial configurations of hard hexagons on the honeycomb lattice for all considered cases,
see Table~\ref{tab01}.
In Table~\ref{tab01}
we also report the energy gap $\Delta$  to the first excited state and the degeneracy of the first excited state.
While in the one-magnon subspace we have $\Delta=J_2-3J_1$, see Eq.~(\ref{202}),
the energy gap in the subspaces $S^z=N/2-n$, $n=2,\ldots,{\cal{N}}/2-1$ agrees with the conjecture 
that for large enough $J_2/J_1 \gtrsim 4$ the first excited states are other localized-magnon states
for which two of the localized magnons are neighbors
(two hard hexagons overlap), see above.
Further evidence for this picture is provided by the value $\Delta=2J_1$ for $S^z=N/4$:
The first excited state with respect to the localized-magnon-crystal state corresponds to three overlapping hard hexagons 
resulting in an increase of energy by $2J_1$
[see also Eq.~(\ref{b04}) in Appendix~B].

\begin{table}
\centering
\caption{${\cal{N}}=12,\,16,\,18,\,24$: 
exact diagonalizations 
[$J_1=1$, 
$J_2=5$ (for ${\cal{N}}=12,\,24$)
and  
$J_2=10$ (for ${\cal{N}}=16,\,18$)]
versus 
counting of the number of hard-hexagon configurations.
DGS is the degeneracy of the ground state,
$\Delta$ is the energy gap,
D1ES is the degeneracy of the first excited state,
\# HHS is the number of configurations of hard hexagons.}
\label{tab01}
\begin{tabular}{|c|c|c|c|c|c|}
\hline
$N$ & $S^z$ & DGS & $\Delta$ & D1ES & \# HHS \\
\hline
 24 &  11   &  12 & 2        &   1  &  12    \\
    &  10   &  48 & 1        &  18  &  48    \\
    &   9   &  76 & 1        & 108  &  76    \\
    &   8   &  45 & 1        & 168  &  45    \\
    &   7   &  12 & 1        &  48  &  12    \\
    &   6   &   2 & 2        &  42  &   2    \\
    &   5   &  12 & 1        &  48  & ---    \\
\hline
 32 &  15   &  16 & 7        &    1 &  16    \\
    &  14   &  96 & 1        &   24 &  96    \\
    &  13   & 272 & 1        &  240 &  272   \\
    &  12   & 376 & 1        &  816 &  376   \\
    &  11   & 240 & 1        & 1104 &  240   \\
    &  10   & --- & 1        &  --- &   72   \\
    &   9   & --- & 1        &  --- &   16   \\
    &   8   & --- & 2        &  --- &    2   \\
\hline
 36 &  17   &  18 & 7        &    1 &   18   \\
    &  16   & 126 & 1        &   27 &  126   \\
    &  15   & 438 & 1        &  324 &  438   \\
    &  14   & 801 & 1        & 1404 &  801   \\
    &  13   & --- & 1        &  --- &  756   \\
    &  12   & --- & 1        &  --- &  348   \\
    &  11   & --- & 1        &  --- &   90   \\
    &  10   & --- & 1        &  --- &   18   \\
    &   9   & --- & 2        &  --- &    2   \\
\hline
 48 &  23   &  24 & 2        &   1 &    24  \\
    &  22   & 240 & 1        &  36 &   240  \\
    &  21   &1304 & 1        & 648 &  1304  \\
    &  20   & --- & ---      & --- &  4212  \\
    &  19   & --- & ---      & --- &  8328  \\
    &  18   & --- & ---      & --- & 10036  \\
    &  17   & --- & ---      & --- &  7176  \\
    &  16   & --- & ---      & --- &  2964  \\
    &  15   & --- & ---      & --- &   752  \\
    &  14   & --- & ---      & --- &   156  \\
    &  13   & --- & ---      & --- &    24  \\
    &  12   & --- & ---      & --- &     2  \\
\hline
\end{tabular}
\end{table}

\begin{figure}
\begin{center}
\includegraphics[clip=on,width=80mm,angle=0]{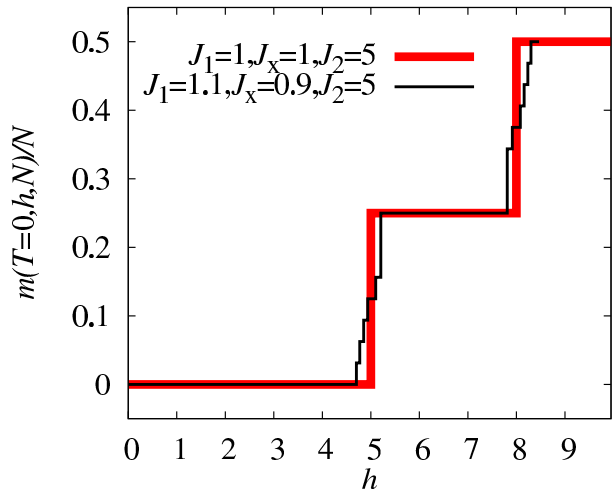}
\caption{(Color online)   Exact-diagonalization results for the zero-temperature magnetization curve
of the honeycomb-lattice bilayer spin-1/2 Heisenberg
antiferromagnet.
The thick solid red line is for the ideal frustration case at $J_1=1$,
$J_2=5$.
Although the data refer to the lattice of $N=32$ sites they do not show finite-size effects.
The magnetization curve has two jumps: 
at $h=h_2=J_2=5$ and $h=h_{\rm{sat}}=J_2+3J_1=8$.
The thin solid black curve shows the zero-temperature magnetization curve for $J_1=1.1$, $J_{\mbox{x}}=0.9$, $J_2=5$ for $N=32$, 
i.e., slightly away from the ideal frustration case,
see also Sec.~\ref{sec4b}.}
\label{f04}
\end{center}
\end{figure}

The zero-temperature magnetization curve is shown by the thick solid red curve in Fig.~\ref{f04}.
The magnetization curve probes the ground-state manifold and it is in a perfect agreement with
the above described picture.
There are two characteristic fields,
$h_2=J_2$
and
$h_{\rm{sat}}=J_2+3J_1$,
at which the ground-state magnetization curve has a jump.
To demonstrate  the robustness of the main features of the magnetization curve against deviations from the ideal frustration case, 
we also show the curve when $J_{\mbox{x}}$ slightly differs from $J_1$.
A more detailed discussion of this issue is then provided in Sec.~\ref{sec4b}.

In the next section we use the established correspondence between the spin model and the hard-hexagon model 
to calculate the thermodynamic properties of the frustrated honeycomb-lattice bilayer quantum Heisenberg antiferromagnet
in the strong-field low-temperature regime.

\section{Hard hexagons on the honeycomb lattice}
\label{sec3}
\setcounter{equation}{0}

The lowest eigenstates in the subspaces with large $S^z$ become ground states for strong magnetic fields.
Thus, 
the energy of these lowest eigenstates in the subspaces with $S^z=N/2-n$, $n=0,1,\ldots,n_{\max}$ in the presence of the field $h$
is
\begin{eqnarray}
\label{301}
E^{{\rm{lm}}}_n(h)=E_{\rm{FM}}-h\frac{N}{2}-\left(\epsilon_1-h\right)n,
\;
\epsilon_1=J_2+3J_1 . \; \;
\end{eqnarray}
At the saturation field, 
i.e., at $h=h_{{\rm{sat}}}=\epsilon_1$, all these energies 
become independent of $n$,
$E^{{\rm{lm}}}_n(h_{{\rm{sat}}})=E_{\rm{FM}}-\epsilon_1 N/2$.
Therefore, the system exhibits  a huge ground-state degeneracy at
$h_{{\rm{sat}}}$ which grows exponentially with the system size $N$:\cite{prb2004,review2}
${\cal{W}}=\sum_{n=0}^{{\cal{N}}/2}g_{{\cal{N}}}(n)\approx\exp(0.218 N)$, see Eq.~(\ref{308}) below.
Here $g_{{\cal{N}}}(n)$ denotes the degeneracy of the ground state for the $2{\cal{N}}$-site frustrated honeycomb-lattice bilayer in the subspace with $S^z=N/2-n$.

Furthermore,
following Refs.~\onlinecite{mike} and \onlinecite{review2},
the contribution of the independent localized-magnon states to the partition function is given by the following formula:
\begin{eqnarray}
\label{302}
Z_{{\rm{lm}}}(T,h,N)=\sum_{n=0}^{\frac{{\cal{N}}}{2}}g_{{\cal{N}}}(n)\exp\left[-\frac{E^{{\rm{lm}}}_n(h)}{T}\right].
\end{eqnarray}
Since $g_{{\cal{N}}}(n)=Z_{{\rm{hc}}}(n,{\cal{N}})$ is the canonical partition function of $n$ hard hexagons on the ${\cal{N}}$-site honeycomb lattice,
Eq.~(\ref{302}) can be rewritten as
\begin{eqnarray}
\label{303}
Z_{{\rm{lm}}}(T,h,N)=\exp\left(-\frac{E_{\rm{FM}}-h\frac{N}{2}}{T}\right) \Xi_{{\rm{hc}}}(T,\mu,{\cal{N}}),
\nonumber\\
\Xi_{{\rm{hc}}}(T,\mu,{\cal{N}})=\sum_{n=0}^{\frac{{\cal{N}}}{2}}Z_{{\rm{hc}}}(n,{\cal{N}})\exp\left(\frac{\mu n}{T}\right),
\mu=\epsilon_1-h. \; \;
\end{eqnarray}
As a result, 
we get the following relations:
\begin{eqnarray}
\label{304}
\frac{F_{{\rm{lm}}}(T,h,N)}{N}
=
\frac{E_{\rm{FM}}}{N}-\frac{h}{2}+\frac{1}{2}\frac{\Omega_{\rm{hc}} (T,\mu,{\cal{N}})}{{\cal{N}}},
\nonumber\\
\Omega_{\rm{hc}} (T,\mu,{\cal{N}})
=
-T\ln\Xi_{\rm{hc}}(T,\mu,{\cal{N}}) \qquad
\end{eqnarray}
for the free energy per site $f(T,h)$,
\begin{eqnarray}
\label{305}
\frac{M_{{\rm{lm}}}(T,h,N)}{N}
=
\frac{1}{2}+\frac{1}{2}\frac{\partial}{\partial\mu}\frac{\Omega_{\rm{hc}}(T,\mu,{\cal{N}})}{{\cal{N}}}
\end{eqnarray}
for the magnetization per site $m(T,h)$,
\begin{eqnarray}
\label{306}
\frac{S_{{\rm{lm}}}(T,h,N)}{N}
=
\frac{1}{2}\frac{S_{{\rm{hc}}}(T,\mu,{\cal{N}})}{{\cal{N}}}
\end{eqnarray}
for the entropy per site $s(T,h)$,
\begin{eqnarray}
\label{307}
\frac{C_{{\rm{lm}}}(T,h,N)}{N}
=
\frac{1}{2}\frac{C_{{\rm{hc}}}(T,\mu,{\cal{N}})}{{\cal{N}}} 
\end{eqnarray}
for the specific heat per site $c(T,h)$.
Note that $h$ and $\mu$ are related by $\mu=h_{\rm{sat}}-h$.
The hard-hexagon quantities in the r.h.s. of these equations depend on the temperature and
the chemical potential only through the activity $z=\exp(\mu/T)$.
That means that for the frustrated quantum spin system at hand all thermodynamic
quantities depend on temperature and magnetic field only via $x=(h_{\rm{sat}}-h)/T=\ln z$,
i.e., a universal behavior emerges in this regime.

To check the formulas for thermodynamic quantities  given in Eqs.~(\ref{304}) -- (\ref{307})
we compare the exact-diagonalization data with the predictions based on the hard-hexagon picture.
We set $J_1=1$, $J_2=5$ and perform exact-diagonalization calculations 
for thermodynamics for the frustrated quantum spin system of $N=24$ sites,\cite{spin} see Fig.~\ref{f03}, 
where the total size of the Hamiltonian matrix is already 
$16\,777\,216 \times 16\,777\,216$.
We also perform the simpler calculations for the corresponding hard-hexagon systems, 
see Appendix~B.

\begin{figure}
\begin{center}
\includegraphics[clip=on,width=80mm,angle=0]{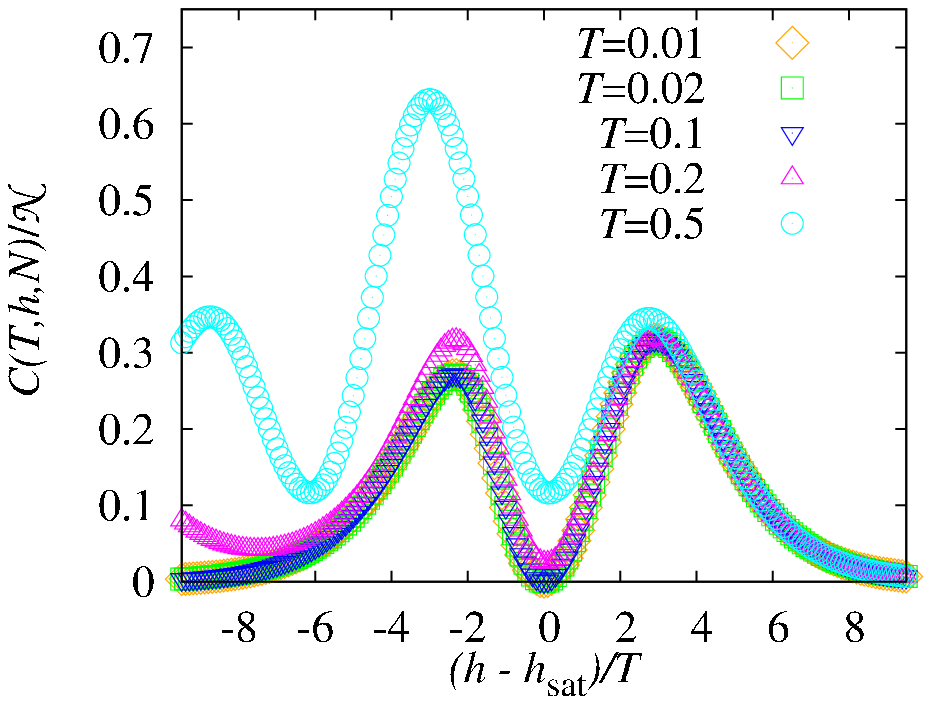}\\
\includegraphics[clip=on,width=80mm,angle=0]{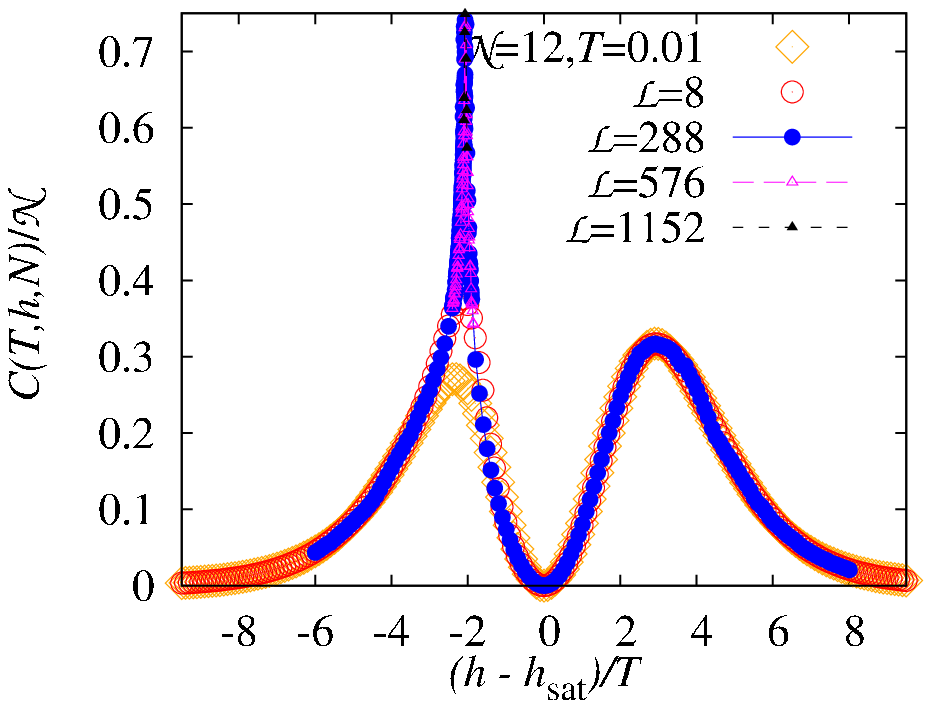}
\caption{(Color online)
Specific heat versus field at low temperatures.
Upper panel:
Exact-diagonalization data for $J_1=1$, $J_2=5$, $N=24$ (${\cal{N}}=12$).
The results for $T=0.01,\,0.02,\,0.1$ are almost indistinguishable.
The results for $T=0.2$ and $T=0.5$ start to deviate from the universal dependence
on $(h-h_{\rm{sat}})/T$.
Lower panel:
Exact-diagonalization data for ${\cal{N}}=12$ (empty diamonds) 
and 
classical Monte Carlo data for ${\cal{N}}={\cal{L}}^2$ with ${\cal{L}}=288,\,576,\,1152$.
Empty circles correspond to the direct calculation for hard hexagons on the honeycomb lattice with ${\cal{N}}=64$.}
\label{f05}
\end{center}
\end{figure}

\begin{figure}
\begin{center}
\includegraphics[clip=on,width=80mm,angle=0]{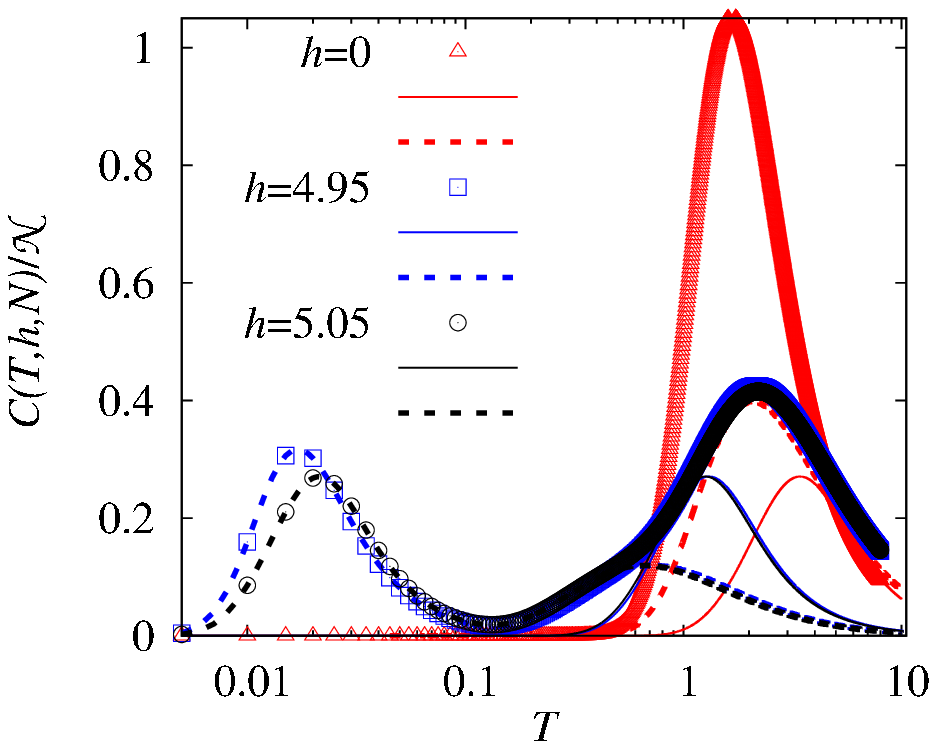}\\
\includegraphics[clip=on,width=80mm,angle=0]{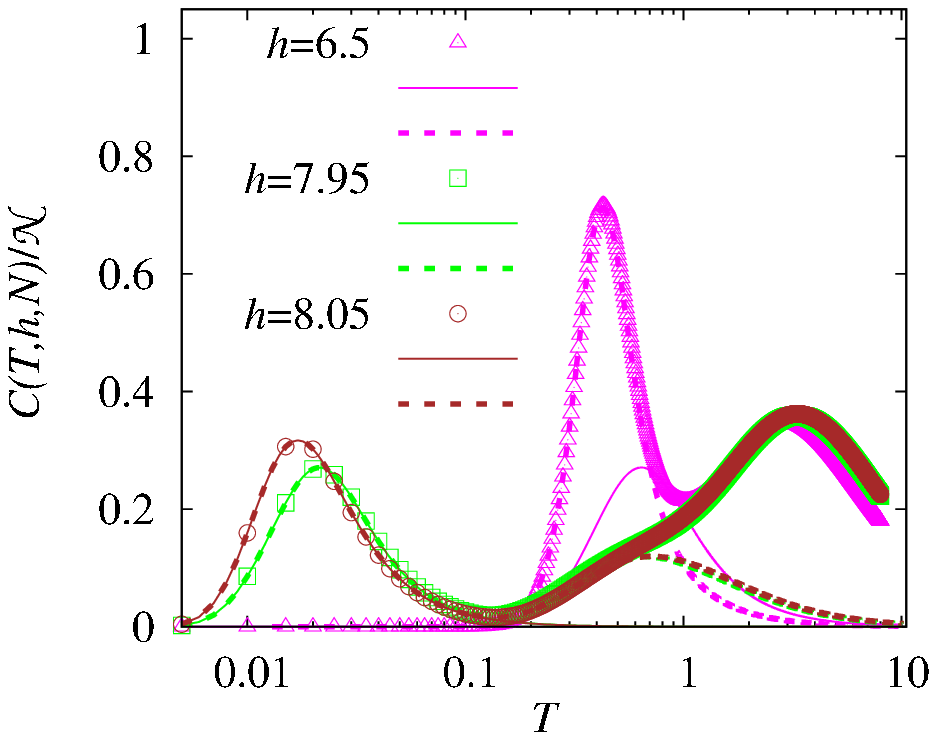}\\
\includegraphics[clip=on,width=80mm,angle=0]{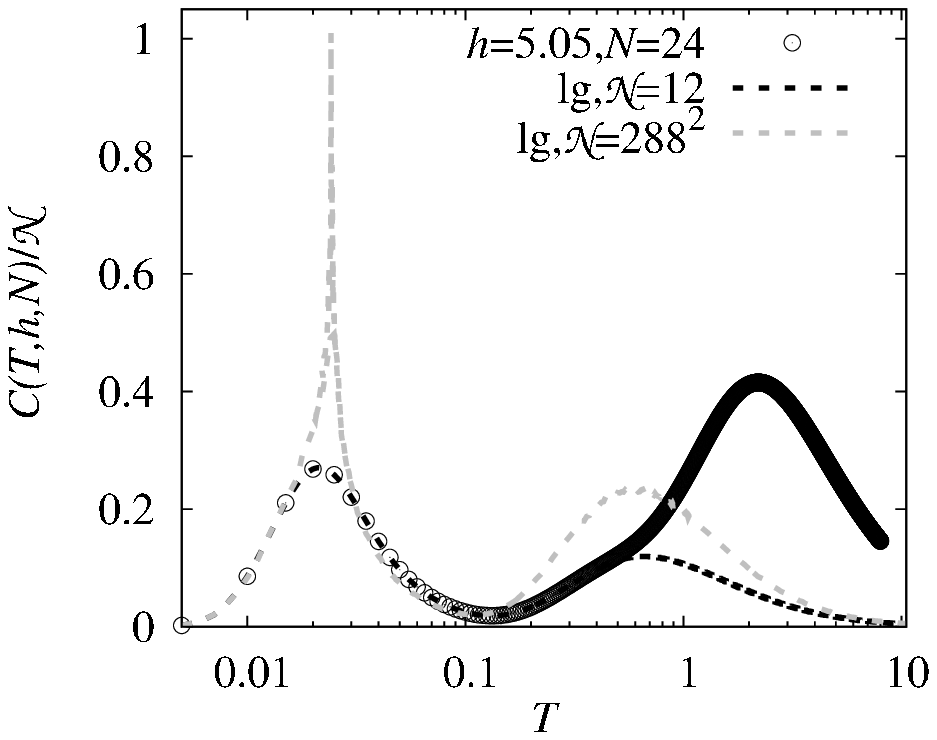}
\caption{(Color online)
Temperature dependence of the specific heat $C(T,h,N)/{\cal{N}}$ for $J_1=1$, $J_2=5$.
Upper panel:
$h=0$ (red),
$h=4.95$ (blue),
and 
$h=5.05$ (black).
Middle panel:
$h=6.5$ (magenta),
$h=7.95$ (green),
and 
$h=8.05$ (brown).
Lower panel:
$h=5.05$.
Exact-diagonalization data (symbols) were obtained for the lattice of $N=24$ sites.
Hard-hexagon predictions (\ref{307}), (\ref{303}) are shown by thin solid lines.
Lattice-gas-model predictions (\ref{402}) are shown by thick dashed lines.
In the lower panel the Monte Carlo data for the lattice-gas model of $288\times 288$ sites are shown by the thick dashed gray line.}
\label{f06}
\end{center}
\end{figure}

Our results for temperature dependences of the specific heat around the saturation are collected in Figs.~\ref{f05} and \ref{f06}.
As can be seen from these plots,
the hard-hexagon picture perfectly reproduces the low-temperature features of the frustrated quantum spin model around the saturation field.
Deviations from the  hard-core-model predictions in the upper panel of Fig.~\ref{f05} become visible only at $T=0.2$.
From the middle panel of Fig.~\ref{f06} one can conclude 
that the temperature profiles for specific heat at $h=7.95$ and $h=8.05$ are well described by the hard-core model again up to about $T=0.1$.


Using the correspondence between the frustrated quantum spin model and the classical hard-core-object lattice-gas model,
we can give a number of predictions for the former model based on the analysis of the latter one.
For example,
we can calculate the ground-state entropy at the saturation field:
\begin{eqnarray}
\label{308}
\frac{S(T\to 0,h=h_{\rm{sat}},N)}{2{\cal{N}}}
=\frac{\ln\Xi_{{\rm{hc}}}(z=1,{\cal{N}})}{2{\cal{N}}}
\approx 0.218.
\end{eqnarray}
This number follows by direct calculations for finite lattices up to ${\cal{N}}=64$ sites.
On the other hand,
for the problem of hard hexagons on a honeycomb lattice
$\kappa=\exp[\ln\Xi_{{\rm{hc}}}(z=1,{\cal{N}})/{\cal{N}}]= 1.546\ldots$
plays 
the same role as 
the hard-square entropy constant
$\kappa=1.503048082\ldots$
for hard squares on the square lattice
or 
the hard-hexagon entropy constant 
$\kappa=1.395485972\ldots$
for hard hexagons on the triangular lattice.\cite{baxter}
Such constants determine the asymptotic growth and are also of interest to combinatorialists.
A more precise value of this constant for hard hexagons on a honeycomb lattice can be found in Ref.~\onlinecite{baxter2}.

The most interesting consequence of the correspondence between the frustrated quantum bilayer and the hard-core lattice gas is the existence of an order-disorder phase transition.
It is generally known
that for the lattice-gas model on the honeycomb lattice with first neighbor exclusion
the  
hard hexagons spontaneously occupy one of two sublattices of the honeycomb lattice
as the activity $z$ exceeds the critical value $z_c=7.92\ldots$, see Ref.~\onlinecite{baxter2}.
In the spin language,
this corresponds to the ordering of the localized magnons as their density increases. 
This occurs at low temperatures just below the saturation field.
For the fixed (small) deviation from the saturation field, 
$h_{\rm{sat}}-h$,
the formula for the critical temperature reads:
\begin{eqnarray}
\label{309}
T_c=\frac{h_{\rm{sat}}-h}{\ln z_c}
\approx 0.48\left(h_{\rm{sat}}-h\right).
\end{eqnarray}
Furthermore,
the critical behavior falls into the universality class of the two-dimensional-Ising-model.\cite{debierre}
That means, the specific heat at $T_c$ (\ref{309}) shows a logarithmic singularity.
Of course, the calculated $T_c$ (\ref{309}) must be small,
otherwise the elaborated effective low-energy theory fails,
see Figs.~\ref{f05} and \ref{f06}.

\section{Beyond independent localized-magnon states}
\label{sec4}
\setcounter{equation}{0}

\subsection{Other localized-magnon states}
\label{sec4a}

Following Ref.~\onlinecite{prb2010},
in addition to the independent localized-magnon states 
(which obey the hard-hexagon rule) 
we may also take into account another class of localized-magnon states which correspond to overlapping hexagon states 
(i.e., they violate the hard-hexagon rule), 
see also our discussion in Sec.~\ref{sec2}.
The corresponding lattice-gas Hamiltonian has the form:
\begin{eqnarray}
\label{401}
{\cal{H}}(\{n_m\})
=
-\mu\sum_{m=1}^{\cal{N}}n_m
+V\sum_{\langle mn\rangle} n_mn_n.
\end{eqnarray}
Here 
$n_m=0,1$ is the occupation number attached to each site $m=1,\ldots,{\cal{N}}$ of the auxiliary honeycomb lattice,
the first (second) sum runs over all sites (nearest-neighbor bonds) of this auxiliary lattice,
and $\mu=h_{\rm{sat}}-h$, $V=J_1$.
The interaction describes the energy increase if two neighboring sites are occupied by hexagons.
In the limit $V\to\infty$ the hard-core rule is restored.

The partition function is given by
\begin{eqnarray}
\label{402}
Z_{{\rm{LM}}}(T,h,N)
=
\exp\left(-\frac{E_{\rm{FM}}-h\frac{N}{2}}{T}\right) \Xi_{{\rm{lg}}}(T,\mu,{\cal{N}}),
\nonumber\\
\Xi_{{\rm{lg}}}(T,\mu,{\cal{N}})
=\sum_{n_1=0,1}\ldots\sum_{n_{\cal{N}}=0,1}
\exp\left[-\frac{{\cal{H}}(\{n_m\})}{T}\right]. 
\end{eqnarray}
Since  $Z_{{\rm{LM}}}$ contains not only the contribution from independent localized-magnon states, 
but also from  overlapping localized-magnon states, 
it is valid in a significantly wider region of magnetic fields and temperatures.

\begin{figure}
\begin{center}
\includegraphics[clip=on,width=80mm,angle=0]{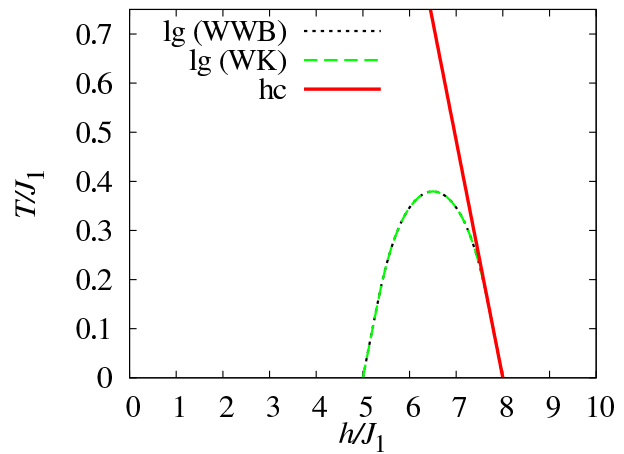}
\caption{(Color online)
Phase diagram of the frustrated honeycomb-lattice bilayer spin-1/2 Heisenberg antiferromagnet
in the plane ``magnetic field -- temperature''.
The coordinates of the highest point of the dome are $h/J_1=J_2/J_1+3/2$ and $T_c/J_1=1/[2\ln (2+\sqrt{3})]\approx 0.380$.
The dome touches the horizontal axis at $h_2/J_1=J_2/J_1$ and $h_{\rm{sat}}/J_1=J_2/J_1+3$.
The black doted and the green dashed lines correspond to the (approximate but very accurate) closed-form expressions 
suggested in Refs.~\onlinecite{honeycomb-ising1,honeycomb-ising2}
(they cannot be distinguished in this graphic representation).}
\label{f07}
\end{center}
\end{figure}

Evidently,
new Ising variables $\sigma_m=2n_m-1$ may be introduced in Eqs.~(\ref{401}) and (\ref{402})
and as a result we face the antiferromagnetic honeycomb-lattice Ising model in a uniform magnetic field:
\begin{eqnarray}
\label{403}
{\cal{H}}
=
{\cal{N}}\left(-\frac{\mu}{2}+\frac{3V}{8}\right)
-\Gamma\sum_{m=1}^{\cal{N}}\sigma_m
+{\cal{J}}\sum_{\langle mn\rangle}\sigma_m\sigma_n,
\nonumber\\
\Gamma=\frac{\mu}{2}-\frac{3V}{4},
\;\;\,
{\cal{J}}=\frac{V}{4}>0.
\end{eqnarray}
The Ising variable $\sigma_m$ acquires two values $\pm 1$,
the nearest-neighbor interaction ${\cal{J}}=J_1/4>0$ is antiferromagnetic,
and the effective magnetic field $\Gamma=(h_{\rm{sat}}-h)/2 -3J_1/4=(J_2+3J_1/2-h)/2$ is zero when $h=J_2+3J_1/2$.
The zero-field case (i.e., $\Gamma=0$) is exactly solvable, 
see Ref.~\onlinecite{jozef-michal} and references therein.
For example, 
the critical temperature is known to be
$T_c/J_1=1/[2\ln (2+\sqrt{3})]\approx 0.380$.
The ground-state antiferromagnetic order in the model (\ref{403}) survives at $T=0$
at small fields 
$\vert\Gamma\vert<3{\cal{J}}$, i.e.,
for $h_2<h<h_{\rm{sat}}$, $h_2=J_2$, $h_{{\rm{sat}}}=J_2+3J_1$.
The antiferromagnetic honeycomb-lattice Ising model in a uniform magnetic field
was a subject of several studies in the past.\cite{honeycomb-ising1,honeycomb-ising2,honeycomb-ising3,honeycomb-ising4}
In particular,
several closed-form expressions for the critical line in the plane ``magnetic field -- temperature''
which are in good agreement with numerical results were obtained,
see Refs.~\onlinecite{honeycomb-ising1,honeycomb-ising2} and also Refs.~\onlinecite{honeycomb-ising3,honeycomb-ising4}.
On the basis of these studies we can construct the phase diagram, see Fig.~\ref{f07}.
Here we have used the two closed-form expressions for the critical line of the antiferromagnetic Ising model in a magnetic field  
suggested in Refs.~\onlinecite{honeycomb-ising1} and \onlinecite{honeycomb-ising2}, 
where both are indistinguishable in the scale used in Fig.~\ref{f07}.
Although the two-dimensional Ising model in a field has not been solved
analytically,
the results of Refs.~\onlinecite{honeycomb-ising1,honeycomb-ising2} are known to be very accurate.\cite{honeycomb-ising2,honeycomb-ising3,honeycomb-ising4}

In Fig.~\ref{f06} 
the temperature profiles for the specific heat in a wide range of magnetic fields
are shown.
The comparison with the exact-diagonalization data demonstrates a clear improvement of the hard-hexagon description
after using the lattice-gas model (\ref{402}).
Furthermore,
in the lower panel of Fig.~\ref{f06} we report classical Monte Carlo data for $h=5.05$ 
[lattice-gas model (\ref{402})]
which shows how the temperature profile $C(T,h,N)/{\cal{N}}$ modifies and develops a singularity
as the system size increases
[see $C(T,h,N)/{\cal{N}}$ for ${\cal{N}}=288^2$ in the lower panel of Fig.~\ref{f06}].

\subsection{Deviation from the ideal flat-band geometry}
\label{sec4b}

Following Ref.~\onlinecite{around},
we can consider an effective low-energy description
when the flat-band conditions are slightly violated and the former flat band acquires a small dispersion.
To this end,
we assume that the intralayer nearest-neighbor interaction $J_1$ and the interlayer frustrating interaction $J_{\mbox{x}}$ are different,
but the difference is small $\vert J_1-J_{\mbox{x}}\vert/J_2\ll 1$.
Then in the strong-field low-temperature regime there are two relevant states at each dimer:
$\vert u\rangle=\vert\uparrow\uparrow\rangle$ 
and 
$\vert d\rangle=(\vert\uparrow\downarrow\rangle - \vert\downarrow\uparrow\rangle)/\sqrt{2}$.
Their energies, 
$\epsilon_u=J_2/4-h$
and
$\epsilon_d=-3J_2/4$,
coincide at $h=h_0=J_2$.
Now the $2^{{\cal{N}}}$-fold degenerate ground-state manifold is splitted by the perturbation,
which consists of the Zeeman term $-(h-h_0)\sum_i s_i^z$
and the interdimer interactions with the coupling constants $J_1$ and $J_{\mbox{x}}$.
The effective Hamiltonian acting in the ground-state manifold can be found perturbatively:\cite{fulde,mila}
\begin{eqnarray}
\label{404}
H_{{\rm{eff}}}=PHP+\ldots,
\end{eqnarray}
where $P=\vert\varphi_0\rangle\langle\varphi_0\vert$ is the projector onto the ground-state manifold,
$\vert\varphi_0\rangle=\prod_{m=1}^{{\cal{N}}}\vert {\rm{v}}\rangle$,
where $\vert {\rm{v}}\rangle$ is either the state $\vert u\rangle$ or the state $\vert d\rangle$.
After some straightforward calculations and introducing the (pseudo)spin-1/2 operators
$T^z=(\vert u\rangle\langle u\vert - \vert d\rangle\langle d\vert)/2$,
$T^+= \vert u\rangle\langle d\vert$,
$T^-= \vert d\rangle\langle u\vert$
at each vertical bond
we arrive at the following result:
\begin{eqnarray}
\label{405}
H_{\rm{eff}}
=
{\cal{N}}\left(-\frac{h}{2}-\frac{J_2}{4}+\frac{3J}{8}\right)
-{\sf{h}}\sum_{m=1}^{\cal{N}}T_m^z
\nonumber\\
+\sum_{\langle mn\rangle}
\left[{\sf{J}}^z T_m^zT_n^z+{\sf{J}}\left(T_m^xT_n^x+T_m^yT_n^y\right)\right],
\nonumber\\
{\sf{h}}=h-J_2-\frac{3J}{2},
J=\frac{J_1+J_{\mbox{x}}}{2},
{\sf{J}}^z=J,
{\sf{J}}=J_1-J_{\mbox{x}}. 
\;\;\;
\end{eqnarray}
The second sum in Eq.~(\ref{405}) runs over all $3{\cal{N}}/2$ nearest-neighbor bonds of the auxiliary honeycomb lattice.
Note that the sign of the coupling constant ${\sf{J}}$ is not important, since the
auxiliary-lattice model (\ref{405}) is bipartite.
Again the effective Hamiltonian (\ref{405}) 
which corresponds to the spin-1/2 $XXZ$ Heisenberg model in a $z$-aligned field on the honeycomb lattice
is much simpler than the initial model and it can be studied further by, e.g., the
quantum Monte Carlo method.\cite{alps}

For the ideal flat-band geometry (ideal frustration case) the effective Hamiltonian (\ref{405}) transforms into the above discussed lattice-gas or Ising models.
To make this evident we have to take into account that 
$J=J_1=V$, 
${\sf{h}}=h-h_{\rm{sat}}+3J/2=-\mu+3V/2$, 
${\sf{J}}^z=J_1=V$, 
${\sf{J}}=0$, 
and replace $T^z$ by $-\sigma/2$:
\begin{eqnarray}
\label{406}
H_{\rm{eff}}
=
{\cal{N}}\left(-\frac{h}{2}-\frac{J_2}{4}+\frac{3V}{8}\right)
\nonumber\\
-\left(\frac{\mu}{2}-\frac{3V}{4}\right)\sum_{m=1}^{\cal{N}}\sigma_m
+\frac{V}{4}\sum_{\langle mn\rangle}\sigma_m\sigma_n
\nonumber\\
=E_{\rm{FM}}-h{\cal{N}}
+{\cal{N}}\left(-\frac{\mu}{2} +\frac{3V}{8}\right)
\nonumber\\
-\left(\frac{\mu}{2}-\frac{3V}{4}\right)\sum_{m=1}^{\cal{N}}\sigma_m
+\frac{V}{4}\sum_{\langle mn\rangle}\sigma_m\sigma_n,
\end{eqnarray}
cf. Eqs.~(\ref{402}) and (\ref{403}).

\begin{figure}
\begin{center}
\includegraphics[clip=on,width=80mm,angle=0]{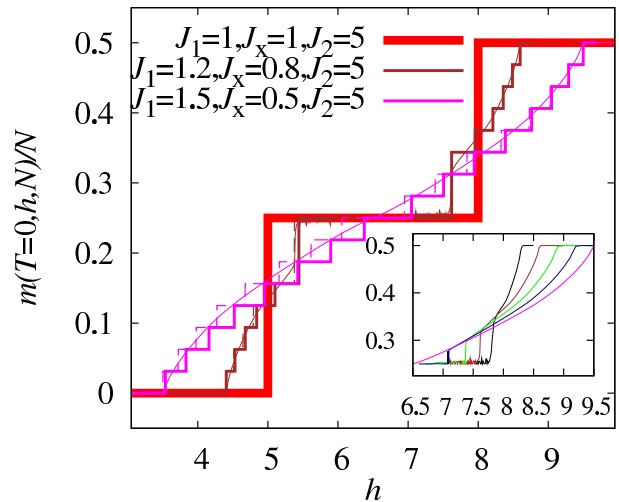}
\caption{(Color online) 
Zero-temperature magnetization curves for the honeycomb-lattice bilayer spin-1/2 Heisenberg antiferromagnet:
Exact diagonalization ($N=32$) and quantum Monte Carlo simulations.
The thick solid red line refers to the ideal frustration case $J_{\mbox{x}}=J_1=1$, $J_2=5$.
Thin lines 
[exact diagonalization for the full model of $N=32$ sites (solid) 
and
for the corresponding effective $XXZ$ model of ${\cal{N}}=16$ sites (dashed)] 
and very thin lines
(quantum Monte Carlo for ${\cal{N}}=2\,304$) 
correspond to deviation from the ideal frustration case,
$J_1=1.2$, $J_{\mbox{x}}=0.8$
and
$J_1=1.5$, $J_{\mbox{x}}=0.5$, 
whereas $J_2=5$.
Note that for $J_1=1.2$, $J_{\mbox{x}}=0.8$
the exact-diagonalization data for the full model and for the corresponding effective model almost coincide.
In the inset we show separately more quantum Monte Carlo data (${\cal{N}}=2\,304$) for 
$J_1-J_{\mbox{x}}=0.2$ (black), 
$J_1-J_{\mbox{x}}=0.4$ (brown), 
$J_1-J_{\mbox{x}}=0.6$ (green), 
$J_1-J_{\mbox{x}}=0.8$ (dark blue), 
and 
$J_1-J_{\mbox{x}}=1$ (magenta).}
\label{f08}
\end{center}
\end{figure}

To illustrate the quality of the effective description,
we compare the results for the ground-state magnetization curve obtained by exact diagonalization
for the initial model of $N=32$ sites
(thin solid curves in Fig.~\ref{f08})
and
for the effective model of ${\cal{N}}=16$ sites
(thin dashed curves in Fig.~\ref{f08}).

It is worth noting the symmetry present in the Hamiltonian (\ref{405}):
If one replaces 
$h=J_2+3J+\delta h$ to $h=J_2- \delta h$
and
all $T_m^z$ to $-T_m^z$
the Hamiltonian (\ref{405}) (up to the constant) remains the same.
This symmetry of the effective model is also present in the exact-diagonalization 
data for the initial model,
if deviations from the flat-band geometry are small,
see the thin solid black curve and the thin solid brown curve in Figs.~\ref{f04} and \ref{f08}.
Moreover, it is also obvious in the lattice-gas Hamiltonian (\ref{401}):
After the replacement 
$\mu=\delta\mu$ to $\mu=3J_1-\delta\mu$ 
and
all $n_m$ to $1-n_m$
the Hamiltonian (\ref{401}) (up to the constant) remains the same.

As can be seen in Fig.~\ref{f08},
the magnetization jumps survive even for moderate deviations from the ideal frustration
case.
The nature of the jump is evident from the effective model (\ref{405}):
It is a spin-flop transition, 
which is present in a two-dimensional Ising-like $XXZ$ Heisenberg antiferromagnet in an external field along the easy axis,
see, e.g., Refs.~\onlinecite{spin-flop-yunoki,spin-flop-selke,spin-flop}.
Note that according to Eq.~(\ref{405})  
the effective easy-axis $XXZ$ model becomes isotropic for $J_1+J_{\mbox{x}}=2(J_1-J_{\mbox{x}})$, 
i.e., the spin-flop transition disappears as increasing the deviation from the ideal frustration case $J_1=J_{\mbox{x}}$.
Although, we are not aware of previous studies of the spin-flop transition for 
the honeycomb-lattice spin-1/2 $XXZ$ model
(and such a study is beyond the scope of the present paper),
we may mention here that the square-lattice case was examined in Ref.~\onlinecite{spin-flop-yunoki}.
In particular, 
one may find there the dependences of the height of the magnetization jump and of the
transition field on the anisotropy.
Furthermore, for temperature effects, see Ref.~\onlinecite{spin-flop-selke}.
Supposing, that for the honeycomb-lattice case the same scenario as for the square-lattice case is valid, 
we may expect that the spin-flop like transition in our model only disappears at the isotropic point $J_1+J_{\mbox{x}}=2(J_1-J_{\mbox{x}})$. 
Our quantum Monte Carlo data shown in the inset of Fig.~\ref{f08} support this conclusion.

Let us complete this section with a general remark on effective models around the ideal flat-band geometry (the ideal frustration case).
Recalling the findings of Ref.~\onlinecite{around}, 
where several localized-magnon systems including the square-kagome model were examined,
we conclude that the effective model around the ideal flat-band geometry essentially depends on the universality class of the localized-magnon system.
(For a comprehensive discussion of the various universality classes of localized-magnon
systems, see Ref.~\onlinecite{univ_class}.)
While for the square-kagome model falling into the monomer universality class we
obtained the (pseudo)spin-1/2 $XXZ$ models with easy-plane anisotropy,\cite{around}
for the considered frustrated honeycomb-lattice bilayer model, 
which belongs to a hard-hexagon universality class,
we get the (pseudo)spin-1/2 $XXZ$ models with easy-axis anisotropy.
Clearly, the magnitude of the Ising terms in the effective Hamiltonian are related to the specific hard-core rules.

\section{Conclusions}
\label{sec5}
\setcounter{equation}{0}

In this paper
we examine 
the low-temperature properties of the frustrated honeycomb-lattice bilayer spin-1/2 Heisenberg antiferromagnet in a magnetic field.
For the considered model,
when the system has local conservation laws,
it is possible to construct a subset of $2^{{\cal{N}}}$ eigenstates
(${\cal{N}}=N/2$)  
of the Hamiltonian and to calculate their contribution to thermodynamics.
For sufficiently strong interlayer coupling, 
these states are low-energy ones for strong and intermediate fields 
and therefore they dominate the thermodynamic properties.

The most interesting features of the studied frustrated quantum spin model are:
The magnetization jumps as well as wide plateaus,
the residual ground-state entropy,
the extra low-temperature peak in the temperature dependence of the specific heat around the saturation,
and the finite-temperature order-disorder phase transition of the two-dimensional Ising-model universality class.
The phase transition occurs just below the saturation field $h_{\rm{sat}}$.
However, for large enough $J_2/J_1$,  
there is a line of phase transitions which occur below $T_c/J_1=1/[2\ln (2+\sqrt{3})]\approx 0.380$
for $h$ in the region between $h_2=J_2$ and $h_{\rm{sat}}=J_2+3J_1$.
Finally,
for deviations from the ideal frustration case we observe for the
isotropic Heisenberg model at hand magnetization jumps which can be understood
as spin-flop like transitions.

There might be some relevance of our study for the magnetic compound
Bi$_3$Mn$_4$O$_{12}$(NO$_3$).
The most intriguing question is:
Can the phase diagram from Fig.~\ref{f07} be observed experimentally?
First, the exchange couplings for Bi$_3$Mn$_4$O$_{12}$(NO$_3$) are still under debate\cite{kandpal} but the relation $J_2/J_1\approx 2$ looks plausible.
In this case the flat band is not the lowest-energy one, see Eq.~(\ref{202}). 
Second, the spin value is $s=3/2$ for this compound 
(each Mn$^{4+}$ ion carries a spin $s=3/2$)
and the localized-magnon effects are less pronounced in comparison with the $s=1/2$
case.
For example, the magnitude ground-state magnetization jump at the saturation is still
${\cal{N}}/2$, but this magnitude is only 1/6 of the saturation value
(in contrast to 1/2 of the saturation value for the $s=1/2$ case). 
Thus, further studies on this compound are needed 
to clarify the relation to the localized-magnon scenario presented in our paper.

\section*{Acknowledgments}

The present study was supported by the Deutsche Forschungsgemeinschaft (project RI615/21-2).
O.~D. acknowledges the kind hospitality of the University of Magdeburg in October-December of 2016.
The work of T.~K. and O.~D. was partially supported by Project FF-30F (No.~0116U001539) from the Ministry of Education and Science of Ukraine.
O.~D. would like to thank the Abdus Salam International Centre for Theoretical Physics (Trieste, Italy) 
for partial support of these studies through the Senior Associate award.

\onecolumngrid

\section*{Appendix A: One-magnon energies (\ref{202})}
\renewcommand{\theequation}{A\arabic{equation}}
\setcounter{equation}{0}

In this appendix,
we present the calculation of the one-magnon energies (\ref{202}).
In the one-magnon subspace,
the Hamiltonian (\ref{201}) can be written in the following form (see Fig.~\ref{f01}):
\begin{eqnarray}
\label{a01}
H=\sum_{m_a=0}^{{\cal{L}}-1}\sum_{m_b=0}^{{\cal{L}}-1}
\left(
J_2 h_{m_a,m_b,1;m_a,m_b,3} +J_2 h_{m_a,m_b,2;m_a,m_b,4}
\right.
\nonumber\\
\left.
+J_1 h_{m_a,m_b,1;m_a,m_b,2}
+J_1 h_{m_a,m_b,1;m_a,m_b,4}
+J_1 h_{m_a,m_b,3;m_a,m_b,4}
+J_1 h_{m_a,m_b,3;m_a,m_b,2}
\right.
\nonumber\\
\left.
+J_1 h_{m_a,m_b,2;m_a,m_b+1,1}
+J_1 h_{m_a,m_b,2;m_a,m_b+1,3}
+J_1 h_{m_a,m_b,4;m_a,m_b+1,3}
+J_1 h_{m_a,m_b,4;m_a,m_b+1,1}
\right.
\nonumber\\
\left.
+J_1 h_{m_a,m_b,2;m_a+1,m_b+1,1}
+J_1 h_{m_a,m_b,2;m_a+1,m_b+1,3}
+J_1 h_{m_a,m_b,4;m_a+1,m_b+1,3}
+J_1 h_{m_a,m_b,4;m_a+1,m_b+1,1}
\right),
\nonumber\\
h_{i;j}
=
\frac{1}{2}\left(s_{i}^-s_{j}^+ + s_{j}^-s_{i}^+\right)
-\frac{1}{2}\left(s_{i}^-s_{i}^+ + s_{j}^-s_{j}^+\right)
+\frac{1}{4}.
\end{eqnarray}
Recall that $N$ is the number of sites,
${\cal{N}}=N/2$ is the number of vertical bonds,
and ${\cal{N}}/2={\cal{L}}^2$ is the number of sites of the triangular lattice which is used here:
The honeycomb-lattice bilayer is viewed as a triangular lattice with four sites in the unit cell.

We introduce the Fourier transformation,
\begin{eqnarray}
\label{a02}
s^+_{m_a,m_b,\alpha}
=
\frac{1}{{\cal{L}}}
\sum_{k_a}\sum_{k_b}\exp\left[{\rm{i}}\left(k_am_a+k_bm_b\right)\right] s^+_{{\bf{k}},\alpha},
\nonumber\\
s^-_{m_a,m_b,\alpha}
=
\frac{1}{{\cal{L}}}
\sum_{k_a}\sum_{k_b}\exp\left[-{\rm{i}}\left(k_am_a+k_bm_b\right)\right] s^-_{{\bf{k}},\alpha},
\nonumber\\
{\bf{k}}=k_a\frac{2}{3a_0}\left(\frac{\sqrt{3}}{2}{\bf{i}}+\frac{1}{2}{\bf{j}}\right)+k_b\frac{2}{3a_0}{\bf{j}},
\nonumber\\
k_a=\frac{2\pi}{{\cal{L}}}z_a,\;z_a=0,1,\ldots,{\cal{L}} -1,
k_b=\frac{2\pi}{{\cal{L}}}z_b,\;z_b=0,1,\ldots,{\cal{L}} -1,
\end{eqnarray}
$a_0$ is the hexagon side length.
After that, Hamiltonian (\ref{a01}) can be cast into
\begin{eqnarray}
\label{a03}
H
=
\frac{{\cal{N}}}{2}\left(\frac{J_2}{2}+3J_1\right)
+
\sum_{{\bf{k}}}
\left(
\begin{array}{cccc}
s^-_{{\bf{k}},1} & s^-_{{\bf{k}},2} & s^-_{{\bf{k}},3} & s^-_{{\bf{k}},4}
\end{array}
\right)
\left(
\begin{array}{cccc}
H_{11} & H_{12} & H_{13} & H_{14} \\
H_{21} & H_{22} & H_{23} & H_{24} \\
H_{31} & H_{32} & H_{33} & H_{34} \\
H_{41} & H_{42} & H_{43} & H_{44} 
\end{array}
\right)
\left(
\begin{array}{c}
s^+_{{\bf{k}},1} \\
s^+_{{\bf{k}},2} \\
s^+_{{\bf{k}},3} \\
s^+_{{\bf{k}},4}
\end{array}
\right)
\end{eqnarray}
with the following matrix ${\bf{H}}$
\begin{eqnarray}
\label{a04}
{\bf{H}}
=
\left(
\begin{array}{cccc}
-\frac{J_2}{2}-3J_1              & \frac{J_1}{2}\gamma_{{\bf{k}}} & \frac{J_2}{2}                    & \frac{J_1}{2}\gamma_{{\bf{k}}} \\
\frac{J_1}{2}\gamma^*_{{\bf{k}}} & -\frac{J_2}{2}-3J_1            & \frac{J_1}{2}\gamma^*_{{\bf{k}}} & \frac{J_2}{2}                  \\
\frac{J_2}{2}                    & \frac{J_1}{2}\gamma_{{\bf{k}}} & -\frac{J_2}{2}-3J_1              & \frac{J_1}{2}\gamma_{{\bf{k}}} \\
\frac{J_1}{2}\gamma^*_{{\bf{k}}} & \frac{J_2}{2}                  & \frac{J_1}{2}\gamma^*_{{\bf{k}}} & -\frac{J_2}{2}-3J_1 
\end{array}
\right),
\nonumber\\
\gamma_{{\bf{k}}}=1+\exp\left(-{\rm{i}}k_b\right)+\exp\left[-{\rm{i}}\left(k_a+k_b\right)\right].
\end{eqnarray}
The eigenvalues of the matrix ${\bf{H}}$ are as follows:
\begin{eqnarray}
\label{a05}
\left\{ 
-J_2-3J_1,
\;\;\;
-J_2-3J_1,
\;\;\;
-3J_1-J_1\vert \gamma_{{\bf{k}}} \vert,
\;\;\;
-3J_1+J_1\vert \gamma_{{\bf{k}}} \vert
\right\},
\nonumber\\
\vert \gamma_{{\bf{k}}} \vert
=\sqrt{3+2\left[\cos k_a +\cos k_b +\cos\left(k_a+k_b\right)\right]}.
\end{eqnarray}
Therefore,
in the one-magnon subspace we have
\begin{eqnarray}
\label{a06}
H
=
E_{\rm{FM}}
+
\sum_{{\bf{k}}}
\sum_{\alpha=1,2,3,4}
\Lambda_{{\bf{k}}}^{(\alpha)}{\sf{s}}_{{\bf{k}},\alpha}^-{\sf{s}}_{{\bf{k}},\alpha}^+,
\nonumber\\
\Lambda_{{\bf{k}}}^{(1)}=\Lambda_{{\bf{k}}}^{(2)}
=-J_2-3J_1,
\;\;\;
\Lambda_{{\bf{k}}}^{(3,4)}
=-3J_1\mp J_1\vert \gamma_{{\bf{k}}} \vert.
\end{eqnarray}

\section*{Appendix B: Finite lattices}
\renewcommand{\theequation}{B\arabic{equation}}
\setcounter{equation}{0}

In this appendix,
we collect some formulas for finite lattices.

Consider hard hexagons on the (periodic) ${\cal{N}}=12$ site honeycomb lattice,
see Fig.~\ref{f03}.
Then
\begin{eqnarray}
\label{b01}
\Xi_{{\rm{hc}}}(T,\mu, 12)
=1 + 12z+ 48z^2+ 76z^3 +45z^4 +12z^5+ 2z^6,
\nonumber\\
z=\exp\left(\frac{\mu}{T}\right),
\;\;\,
\mu=h_{\rm{sat}}-h,
\end{eqnarray}
see Table~\ref{tab01}.
All thermodynamic quantities follow from Eq.~(\ref{b01}) according to standard prescriptions of statistical mechanics.

Next,
consider the lattice-gas model with finite nearest-neighbor repulsion on the (periodic) ${\cal{N}}=12$ site honeycomb lattice,
see Fig.~\ref{f03}.
Then
\begin{eqnarray}
\label{b02}
\Xi_{{\rm{lg}}}(T,\mu, 12)
=
\sum_{n_1=0,1} \ldots \sum_{n_{12}=0,1}
\exp
\left[
\frac{\mu}{T}\sum_{m=1}^{12} n_m 
\right.
\nonumber\\
\left.
-\frac{V}{T}
\left(
n_5n_1+n_1n_2+n_2n_3+n_4n_2+n_3n_5+n_9n_4+n_4n_6+n_5n_7+n_6n_7
\right.
\right.
\nonumber\\
\left.
\left.
+n_8n_6+n_7n_9+n_{12}n_8+n_8n_{10}+n_9n_{11}+n_{10}n_{11}+n_1n_{10}+n_{11}n_{12}+n_{12}n_3
\right)
\right]
\nonumber\\
=
\sum_{n_1=0,1} \ldots \sum_{n_{12}=0,1}
z^{\sum_{m=1}^{12} n_m}
\exp\left[
-\frac{V}{T}
\left(
n_5n_1+n_1n_2+n_2n_3+n_4n_2+n_3n_5+n_9n_4+n_4n_6+n_5n_7+n_6n_7
\right.
\right.
\nonumber\\
\left.
\left.
+n_8n_6+n_7n_9+n_{12}n_8+n_8n_{10}+n_9n_{11}+n_{10}n_{11}+n_1n_{10}+n_{11}n_{12}+n_{12}n_3
\right)
\right],
\nonumber\\
\mu=h_{\rm{sat}}-h,
\;\;\,
V=J_1,
\;\;\,
z=\exp\left(\frac{\mu}{T}\right).
\end{eqnarray}
The partition function (\ref{b02}) contains 4096 terms and can be easily calculated.
All thermodynamic quantities follow from Eq.~(\ref{b02}) according to standard prescriptions of statistical mechanics.

Clearly, Eq.~(\ref{b02}) implies a specific numbering of sites in Fig.~\ref{f03}.
However, it can be rewritten in the form that does not depend on the site numbering
[as Eq.~(\ref{b01})].
If we introduce the function $g(k_1,k_2)$ with 
the integer $k_1=0,\ldots, 12$ to count the number of occupied sites
and 
the integer $k_2=0,\ldots,18$ to count the number of bonds which connect the occupied sites,
Eq.~(\ref{b02}) can be cast into
\begin{eqnarray}
\label{b03}
\Xi_{{\rm{lg}}}(T,\mu, 12)
=
\sum_{k_1=0}^{12}\sum_{k_2=0}^{18} g(k_1,k_2) z^{k_1} \exp\left({-\frac{V}{T}k_2}\right).
\end{eqnarray}
The only non-zero values of the function $g(k_1,k_2)$ are as follows:
\begin{eqnarray}
\label{b04}
g(0,0)=1;
\;\;\;
g(1,0)=12;
\;\;\;
g(2,0)=48, g(2,1)=18;
\;\;\;
g(3,0)=76, g(3,1)=108, g(3,2)=36;
\nonumber\\
g(4,0)=45, g(4,1)=168, g(4,2)=207, g(4,3)=72, g(4,4)=3;
\nonumber\\
g(5,0)=12, g(5,1)=48, g(5,2)=276, g(5,3)=276, g(5,4)=168, g(5,5)=12;
\nonumber\\
g(6,0)=2, g(6,1)=0, g(6,2)=42, g(6,3)=212, g(6,4)=342, g(6,5)=264, g(6,6)=62;
\nonumber\\
g(7,3)=12, g(7,4)=18, g(7,5)=276, g(7,6)=276, g(7,7)=168, g(7,8)=12;
\nonumber\\
g(8,6)=45, g(8,7)=168, g(8,8)=207, g(8,9)=72, g(8,10)=3;
\nonumber\\
g(9,9)=76, g(9,10)=108, g(9,11)=36;
\;\;\;
g(10,12)=48, g(10,13)=189;
\;\;\;
g(11,15)=12;
\;\;\;
g(12,18)=1.
\end{eqnarray}

This representation allows to clarify the degeneracy of the first excited state reported in Table~\ref{tab01}:
According to the elaborated picture it is given by the value of $g(k_1,1)$, $k_1=2,3,4,5$.
Furthermore, 
$g(6,1)=0$, 
i.e., one cannot place 6 occupied sites on the 12-site lattice in Fig.~\ref{f03} to have only 1 bond connecting the occupied sites.
The smallest number of bonds connecting occupied sites is 2 and $g(6,2)=42$:
This explains the value of the energy gap $\Delta=2$ and the 42-fold degeneracy of the first excited state.

\end{document}